\documentstyle[aps,multicol,epsf]{revtex}

\begin{document}

\title{Localized states in 2D semiconductors doped with magnetic
impurities in quantizing magnetic field }
\author{$^1$P. Dahan, $^2$V. Fleurov,  $^4$ K. Kikoin, and $^{1,3,5}$
I. D. Vagner\\
$^1$P.E.R.I. Physics and Engineering Research Institute\\ Ruppin
Academic Center, Emek-Hefer 40250, Israel.\\ $^2$Beverly and
Raymond Sackler Faculty of Exact Sciences\\ School of Physics and
Astronomy\\ Tel Aviv University Tel Aviv 69978, Israel.\\
$^3$Grenoble High Magnetic Field Laboratory (CNRS-MPI,FKF) \\
166X,F-38042, Grenoble, Cedex9, France.\\ $^4$Ben-Gurion
University of the Negev\\ Beer-Sheva 84105, Israel\\
$^5$Department of Communication Engineering,\\ Holon Academic
Institute of Technology, POB305, Holon 58102 Israel.}
\maketitle

\begin{abstract}
A theory of magnetic impurities in a 2D electron gas quantized by
a strong magnetic field is formulated in terms of Friedel-Anderson
theory of resonance impurity scattering. It is shown that this
scattering results in an appearance of bound Landau states with
zero angular moment between the Landau subbands. The resonance
scattering is spin selective, and it results in a strong spin
polarization of Landau states, as well as in a noticeable magnetic
field dependence of the $g$ factor and the crystal field splitting
of the impurity $d$ levels.
\end{abstract}

\section{Introduction}

Unique properties of two dimensional electron systems (2DES) in
strong magnetic fields together with their potential applications
in microelectronics put these systems among the hottest topics in
the studies of strongly correlated electron systems. The
fractional quantum Hall effect is the brightest manifestation of
unusual quantum statistics of 2DES (see, e.g., Ref.
\onlinecite{Das}. Yet, even in the case of an integer occupation
an extremely rich variety of excitations  can be observed in these
systems. Various types of excitons \cite{Lelo80,BIE81,KH84} and
collective topological excitations (skyrmions)
\cite{Sondhi93,BMV96} arise due to interplay between orbital and
spin degrees of freedom.

Impurity scattering is responsible for broadening of Landau levels
and creation of localized states in a 2DES. The genesis of the
bound states due to a short-range potential scattering was
described in Refs. \onlinecite{Avishai93,Avishai97}. However, the
role of magnetic impurities in formation of the excitation
spectrum in 2DES was not discussed thoroughly as yet. Meanwhile,
the recent experimental achievements in doping semiconductors with
magnetic impurities in confined geometries are significant. A high
enough concentration of transition metal (TM) impurities is
inserted in heterostructures formed by several III-V and II-VI
semiconductor compounds \cite{Dietl}. For example, in II-VI
heterostructures (Cd,Mn)Te/(Cd,Mg)TE planar concentration of Mn is
comparable with the 2D electron concentration (see, e.g.,
\cite{Potem99,Potem00}). Apparently, in such cases TM is the
dominant impurity that predetermines the filling of the gaps
between the Landau levels. To describe modifications of the energy
spectrum of a 2DES one should first consider the problem of an
isolated magnetic impurity in a 2D electron gas in a strong
quantizing magnetic field. This problem is solved in the present
paper.

Our study of magnetic impurities in a 2DES is based on two
existing approaches: (i) we generalize the theory of TM impurities
developed  for the bulk semiconductors and summarized in Refs.
\onlinecite{Zung86a,KF94}; (ii) we show that the behavior of TM
impurities has much in common with the properties of isoelectronic
point-like impurities, and in a sense our theory is a
generalization of the approach formulated in Refs.
\onlinecite{Avishai93,Avishai97} for the case of {\em spin
selective} resonance impurity scattering that results in an
appearance of bound Landau states between the bare Landau levels.
We find also that the interplay between the orbital and spin
degrees of freedom results in a significant modification of the
$g$ factors both of the impurity $d$ electrons and electrons on
the bound Landau levels.

\section{Impurity levels and bound states in 2DES}

Perturbation inserted by TM impurities in a magnetic field
quantized energy spectrum of electrons in semiconductors can be
described by a Friedel-Anderson resonance scattering model (see,
e.g., \cite{KF94}). According to this model, TM impurity
introduces its $3d-$level either in the semiconductor energy gap
or in the nearest valence or conduction band (depending on the
atomic number of the impurity). As a result, the scattering
amplitude due to the local impurity potential acquires a strong
energy dependence, characteristic of the Friedel-type resonance
scattering. A quantizing magnetic field makes the electron motion
finite in a plane perpendicular to the field direction. Only those
electron orbitals are perturbed by the impurity potential, which
envelop the defect cell. Therefore, the degeneracy of each Landau
level in a 2DES is partially lifted by the impurity scattering. A
similar problem was discussed earlier for the case of potential
scatterers in a quantizing magnetic field (see, e.g.,
\cite{Lelo80,Avishai93,Avishai97,Demd65}. Here we study the case
of resonance scattering in the 2DES, which is formed, e.g in
GaAs/GaAlAs semiconductor heterostructures.

The calculation starts from the impurity Hamiltonian in a magnetic
field $B$ parallel to the $z$-axis,
\begin{equation}
H_{e}=H_0 +V_d ({\bf r}-{\bf R}_{0})
\label{HAM1}
\end{equation}
where
\begin{equation}
H_{0} = \frac{1}{2m^*} \left( {\bf P} + \frac{e}{c}{\bf A} \right)
^{2}+V(z)
\end{equation}
describes the motion of an electron in the conduction band with
the effective mass $m^*$, confined in the $z$ direction by the
potential $V(z)$. $V_d({\bf r}-{\bf R}_{0})$ is the substitutional
impurity potential at a site ${\bf R}_{0}$. The effect of the
periodic lattice potential is taken into account in the effective
mass approximation. It may be applied since we are interested in
the properties of magnetically quantized states near the bottom of
the lowest conduction band. The behavior of the impurity wave
functions and the positions of the impurity levels are
predetermined by the singularities in the spectrum of these
states. Therefore, we neglect in our calculations the
contributions of the higher conduction bands as well as those of
the valence bands. Then the band wave functions assume the form
\begin{equation}
\Psi_{\lambda,j}({\bf r})=\Phi_{\lambda}(\rho,\phi)\chi_j(z).
\label{BWF}
\end{equation}
Here $\lambda,j$ are the quantum numbers describing the finite
electron motion in the $xy$-plane and in the $z$-direction,
respectively. $\chi_j(z)$ are eigenfunctions of the confining
potential $V(z)$. Only one state of this confining potential
closest to the electron Fermi level will be taken into account in
what follows. The corresponding index $j$ will be suppressed
below. It is also convenient for our purposes to describe the
electron motion in the $xy$-plane by means of the cylindrical
coordinates $\rho,\ \varphi$. We choose the cylindrical gauge for
the vector potential ${\bf A} = \displaystyle{( - \frac{B}{2} y,
\frac{B}{2} x, 0)}$. Then $\lambda = nm$ where $n$ is the Landau
level index and $m$ is the orbital quantum number enumerating the
states within a given Landau level. The planar component of the
wave function (\ref{BWF}) has the form
\begin{equation}
\Phi _{nm}(\rho ,\phi ) = C_{nm}\xi ^{|m|/2}L_{n + \frac{m -
|m|}{2}}^{|m|} \left( \xi \right) e^{-im\varphi} e^{-\frac{\xi
}{2}}, \label{LWF}
\end{equation}
\[
\rho = l_B\sqrt{2\xi },\;\;\;l_B^2 = \hbar c/eB,
\]
\[
C_{nm} = \sqrt{\frac{(n+\frac{m-|m|}{2})!}{2\pi l_B^2
(n+\frac{m+|m|}{2})!}}\ \ \ \ (n\geq \pm m\geq -\infty ).
\]
Here $E_n= \displaystyle \frac{\hbar^2}{2m^\ast}
\frac{2n+1}{l_B^2}$ is the energy of the $n$-th Landau level,
$L_{n + \frac{m-|m|}{2}}^{|m|} \left( \xi \right) $ is Laguerre
polynomial.

In accordance with the general scheme of the resonance model
\cite{KF94}, the wave function of an electron, localized in a
discrete impurity level, is represented by an expansion
\begin{equation}
\Phi_{i\gamma \mu } = F_d^{\gamma \mu } \varphi _{\gamma \mu} +
\sum_\lambda F_\lambda^{\gamma \mu } \varphi_\lambda.
\label{IWF}
\end{equation}
The atomic $d$-orbital $\varphi_{\gamma\mu}$ forms the "core" of
the impurity wave function that retains its 3D character, because
its radius $r_d$ is small in comparison with the width of the well
$V(z)$ responsible for the confinement in the $z$ direction. Here
$\gamma = e,\ t_2$ determines the irreducible representation of
the crystalline point group for the $d$ states, $\mu$ enumerates
lines of these irreducible representations. The orbital effect of
the magnetic field is negligible on the atomic scale. The crystal
field splitting of the energy levels,
\begin{equation}
\varepsilon_\gamma = \varepsilon _{d} + \langle \gamma \mu |W|
\gamma \mu \rangle,
\end{equation}
is the only effect of the crystalline environment. Here
\[
W({\bf r}-{\bf R}_0) = \sum_{j\neq 0}U_h({\bf r}-{\bf R}_j)
\]
is the crystal field of the neighboring host ions in the lattice
sites ${\bf R}_j$ acting on the impurity $d$-electrons. The "tail"
$$\Phi_{b\gamma\mu} =\sum_\lambda F_\lambda^{\gamma \mu }
\varphi_\lambda $$
of the impurity wave function (\ref{IWF}) is a superposition of
the wave functions $\varphi_\lambda$ with $\lambda = nm$, which
are obtained by orthogonalizing the functions (\ref{BWF}) to the
$d$ states. This tail falls down at large distances,
$\Phi_{b\gamma\mu}(r) \sim r^{-1} \exp(-\kappa_\gamma r)$. A rough
estimate of the localization parameter is $\kappa_\gamma \approx
\hbar^{-1} \sqrt{2m_e|E_{i\gamma}|}$ where $|E_{i\gamma}|$ is the
depth of the impurity level relative to the nearest Landau
subband. A detailed discussion of the asymptotic behavior of the
impurity wave functions for bulk semiconductors can be found in
the book \cite{KF94}. Due to the obvious symmetry of the problem
it will be convenient for the future analysis to choose the origin
of the system of coordinate coinciding with the impurity.

Substitution of the wave function (\ref{IWF}) into the
Schr\"odinger equation with the Hamiltonian (\ref{HAM1}) results
in the system of equations for the expansion coefficients
$\{F_\lambda^{\gamma \mu},F_d^{\gamma \mu }\}$,
\begin{eqnarray}
\left( \varepsilon_\gamma -E\right) F_{d}^{\gamma \mu} +
\sum_{\lambda}\langle \gamma \mu |V_d|\lambda\rangle
F_{\lambda}^{\gamma\mu }=0,\nonumber\\ \label{SM1}\\ \left(
E_\lambda - E\right) F_\lambda^{\gamma \mu } + \langle \lambda
|V_d|\gamma \mu \rangle F_d^{\gamma \mu } + \sum_{\lambda'}
\langle \lambda ^{\prime} |V_d|\lambda \rangle
F_{\lambda`}^{\gamma \mu } = 0 \nonumber
\end{eqnarray}
($E_\lambda \equiv E_n$). The system of equations (\ref{SM1}) may
be represented in the matrix form (see, e.g., \cite{FK86})
\begin{equation}
\left(
\begin{array}{cc}
{\sf B} & -{\sf V}^T \\
-{\sf V} & {\sf D}
\end{array}
\right) \left(
\begin{array}{c}
{\sf b} \\
{\sf d}
\end{array}
\right) =0
\label{M}
\end{equation}
where
\[
({\sf b},{\sf d}) = (F_\lambda^{\gamma \mu },F_d^{\gamma \mu })
\]
and $T$ denotes the transposition of the matrix. The elements of
the matrices ${\sf B,\ D,\ V}$ are
\begin{eqnarray}
B_{\lambda,\lambda^\prime}=\left( E-E_\lambda\right) \delta
_{\lambda ,\lambda^\prime} - \langle \lambda |V_d|
\lambda'\rangle,\nonumber\\ \nonumber\\ D_{\gamma\mu,
\gamma^\prime \mu^\prime} = \left( E-\varepsilon_\gamma \right)
\delta _{\gamma \gamma^\prime}\delta _{\mu \mu^{\prime }},
\label{D1}\\ \nonumber\\ V_{\gamma\mu,\lambda} = \langle \gamma
\mu |V_d|\lambda\rangle, \nonumber
\end{eqnarray}
respectively. The energy levels are determined by the secular
equation for the system (\ref{M})
\begin{equation}
\det {\sf M} \equiv \det \left( {\sf D}(E)-{\sf VB^{-1}}(E){\sf
V}^T\right) = 0. \label{M3}
\end{equation}
It follows from Eq. (\ref{M3}) that the energies of the impurity
bound states are determined by the equation
\begin{equation}
E - \varepsilon_\gamma - M_\gamma(E) = 0,
\label{DM3}
\end{equation}
where
\begin{equation}
M_{\gamma }(E) = \sum_{\lambda \lambda^\prime} \frac{\langle \gamma \mu
|V_d|\lambda \rangle \langle \lambda |{\sf Q}^{-1}|\lambda^\prime \rangle
\langle \lambda^\prime |V_d| \gamma \mu \rangle }{ E - E_{\lambda}},
\label{SEN}
\end{equation}
\[
{\sf Q = 1-VG},~{\sf G}=({\sf 1}\cdot E - {\sf H}_{0})^{-1},
\]
${\sf 1}$ is the unit matrix.

Eq. (\ref{DM3}) is the basic equation of the theory. It describes
the renormalization of both the impurity $d$-level $E_{i\gamma}$
and of the Landau bands $E_\lambda$. The self energy part
$M_{\gamma}(E)$ contains information about the potential
scattering \cite{FK86}. In particular, the zeros of the matrix
${\sf Q}(E)$ determine the Landau levels modified by the short
range impurity potential $\langle \lambda |V_d| \lambda^\prime
\rangle = V$. Since $V_d$ is nonzero mostly within the impurity
crystalline cell, the dependence of its matrix elements on the
indices of the Landau states is very weak, and we neglect it for
the sake of simplicity. Just this very type of the potential is
used in the theory of the point-like impurity scattering
\cite{Avishai93,Avishai97}. This potential contains a bound
$d$-level so it is responsible both for the resonance and
potential impurity scattering (see \cite{FK86} for more details).
The sign of the potential scattering amplitude depends on specific
characteristics of the substitutional impurity atom relative to
those of the substituted host atom. It can be either negative or
positive.

Bound impurity states in case of a potential scattering with $V <
0$ were investigated in Refs. \cite{Avishai93,Avishai97}. In order
to use the results of those studies it is convenient to represent
the matrix ${\sf M = V G Q^{-1} V}$ in a more symmetric form
\[
{\sf M=V{\tilde G}V}
\]
Here ${\sf \tilde G}(E)$ is the Green function of a single short
range impurity problem found in \cite{Avishai93}. Now the matrix
element $M_\gamma$ (\ref{SEN}) acquires the form
\begin{equation}
M_\gamma (E_{i\gamma }) = \sum_\beta\frac{\langle \gamma \mu |V_d|
\beta \rangle \langle \beta|V_d|\gamma \mu \rangle} {E_{i\gamma} -
E_\beta}. \label{SE}
\end{equation}
where $|\beta \rangle$ stand for the eigenfunctions of magnetic
field quantized electrons in the local potential $V_d$,
\[
|\beta \rangle ={\sf Q}^{-1}|\lambda \rangle.
\]

The most important property of this solution is that all the
states $|nm \rangle$ with nonzero angular momenta ($m \neq 0$) are
left intact by the short range potential $V_d$ because they have
nodes at the impurity site. Thus, $E_\beta = E_\lambda$ for the
states with $m\neq 0$ \cite{Lelo80,Avishai93}. This is true also
for the resonant scattering potential, although in this case the
statement is only approximate. The matrix elements
\[
\langle nm|V_d|\gamma \mu \rangle \sim \left( \frac{\rho_d }{l_B}
\right) ^{|m|}\ll 1
\]
for $m\neq 0$, since the typical localization radius $\rho_d$ of
the impurity d-functions is much smaller than the magnetic length
$l_B$. It means that only one d-orbital, $|e1\rangle \propto
|r^{2}-3z^{2}\rangle \sim Y_{20}$, may be strongly hybridized with
states with $m = 0$. Thus, both potential and resonance components
of the TM impurity act on the same cylindrically symmetric states,
meaning that only the states with $m=0$ enter the self energy
(\ref{SE}). We use the Green function in Eq. (\ref{SE}) in the
form
\begin{equation}
\tilde G(\xi, z;\xi ^{\prime },z^\prime) = \sum_n
\frac{\varphi_b(\xi;E_{n}) \chi(z)\varphi_b^{\ast }(\xi ^{\prime
};E_{n}) \chi^\ast(z^{\prime })}{E-E_{bn}}. \label{2DGF}
\end{equation}
where only the bound states with $m=0$, belonging to the relevant
quantum level $E_j$ of the confinement potential $V(z)$, are
retained. The energy spectrum and the wavefunctions of these
states were calculated in ref. \cite{Avishai93}. Several results
of this paper, which we need for the further discussion, are
presented below.

The wavefunctions $\varphi_b(\xi;E_{bn})$ for the bound Landau
states with $m=0$ have the following asymptotic at large $\xi$,
\begin{equation}
\varphi_b(\xi ,E_{bn}) = \frac{\Gamma(\frac{1}{2} -
\alpha_n)}{\sqrt{2\pi\psi(\frac{1}{2} - \alpha_n))}}
\frac{W_{\alpha_n,0}(\xi)}{l_B\xi^\frac{1}{2}}.
\label{BWF2}
\end{equation}
Here $\alpha_n = \frac{1}{2}\left(1 - \epsilon_{bn} l_B^2\right)$,
$\epsilon_{bn} = 2m_\perp E_{bn}/\hbar ^{2}$,
$W_{\alpha_n,0}(\xi)$ is the Whittaker function, $\psi(\alpha_n)$
is the digamma function. The energy levels $\epsilon_{bn}$, split
from the corresponding Landau levels $\epsilon _{n0}$, are
described by the following equation,
\begin{equation}
\psi(\alpha_n) + \ln \frac{2}{|\epsilon_i|l_B^2} = 0. \label{EWF}
\end{equation}
where $E_{i}$ is the energy level of an electron, bound by the
impurity attractive potential $V$ at $B=0$.

The first bound state appears below the bottom of the conduction
band. Its energy is given by the equation
\begin{equation}
\epsilon_{b0} = \epsilon_{i}-\frac{1}{6\epsilon_{i}l_B^4},
\label{ASY}
\end{equation}
which follows from Eq. (\ref{EWF}), provided the energy level
$E_{i}$ is deep enough (or the magnetic field is weak),
$|\varepsilon_{bi}| \gg l_B^{-2}$. Other discrete levels in this
case are slightly shifted Landau levels. It follows from
(\ref{EWF}) that
\begin{equation}
\epsilon_{bn} = \epsilon_{n-1,0} + 2l_B^{-2} \left | \ln
\frac{2e^{\psi(\alpha_n)}} {|\varepsilon_i|l_B^2} \right|^{-1}.
\label{EWWF}
\end{equation}
With an increase of the magnetic field the impurity level
$\epsilon_n$ moves upward towards $\epsilon_{n0}$.

The positions of the Landau bound states induced by the resonance
scattering component of the impurity potential are determined by
the self energy part
\begin{equation}
M_{e}(E) = \sum_{n}{\frac{ |V_{en}|^2}{E - E_{bn}}}. \label{SE1}
\end{equation}
Therefore, Eq. (\ref{DM3}) becomes
\begin{equation}
E_{ie\sigma} - \varepsilon_{e\sigma} = M_{e}(E_{ie\sigma})
\label{DM4}
\end{equation}
(we restored the $d$ - electron spin indices for the reasons,
which will be explained in the next section). Here the matrix
elements
\begin{equation}
V_{en} = \int d^3r~ \varphi_{e1 }({\bf r}) V_d({\bf r})
\varphi_b(\xi, E_{bn})\chi_{0}(z) \label{int}
\end{equation}
describe the hybridization of the atomic and bound Landau states.
They are magnetic field dependent (see Section 4 for further
discussion).

Two lowest solutions of Eq. (\ref{DM4}) arise below the first
Landau level. To illustrate the mechanism of the renormalization
we neglect all the Landau levels in this equation except for the
lowest one, described by Eq. (\ref{ASY}). Then two first discrete
levels follow from the simplified equation,
\begin{equation}
(E_{i\sigma}^{(b,a)} - \varepsilon_{e\sigma})
(E^{(b,a)}_{i\sigma}-E_{b0}) = |V_{eb}|^2. \label{BA}
\end{equation}
If the resonance $de1$-level arises deep below the quantized band
level $E_{n=0}$, then the bonding state $E_{i\sigma}^{(b)}$ is the
renormalized impurity $e1$-level and the antibonding state
$E_{i}^{(a)}$ is the former localized level $E_{b0}$ shifted
upwards. In this case the resonance scattering is weak, and it
cancels partially the contribution of the short range potential
$V_d$. If the resonance level appears above $E_{b0}$, then the
(now bonding) level $E_{i}^{(b)}$ deepens in comparison with the
level (\ref{ASY}). The qualitative graphical solution of Eq.
(\ref{DM4}) is presented in Fig. 1.

Fig. 1a exhibits a graphical solution of Eq. (\ref{DM4}) in the
general case. Here two lowest solutions correspond to the states
$E_{i}^{(b,a)}$, and all the remaining solutions represent the
states with $m=0$ split from the degenerate Landau levels
$E_{n0}$. In fact, all these states are solutions of Eqs.
(\ref{EWF}) or (\ref{EWWF}) shifted upward. Fig. 1b illustrates
graphical solution of simplified secular equation (\ref{BA}).

It is known \cite{KF94} that the resonance level $E_e$ of TM
impurities in a neutral state with configuration $3d^n$ always
arises below the bottom of the conduction band. However, in the
case of charged impurity states $3d^{n+1}$ the bare $e$ - level
may appear very close to the bottom of the conduction band. For
example, vanadium impurity V$^{2+}$ in GaAs possesses just this
kind of spectrum \cite{Zung86}. In some cases (e.g., Cr in GaAs)
the e-state of the charged impurity may appear above the bottom of
the conduction band \cite{KF94}, and this is the case of a strong
resonance scattering, leading to a shift of the levels $E_{bn}$
downward (dashed line in Fig. 1a). A further discussion of
possible experimental realizations of the strong resonance
impurity scattering is presented in the following sections.

Since $3d$-impurity interacts only with the band orbitals in its
nearest vicinity, one should take into account the scatter in the
positions of the Landau levels in different parts of the sample
due to long range fluctuations of the local fields. This scatter
is described by a Gaussian distribution, so the split-off levels
with $m=0$ form a replica of this distribution. As a result, in
real samples one should observe the density of states presented in
Fig.2.

The removal of the symmetry selection rules for the $e$ states
also influences significantly the crystal field splitting
$\Delta_{CF} = E_{it_2} - E_{ie}$ of the impurity levels. For
example, in the particular case of the charged V$^{2+}$ impurity
in GaAs the level $E_{ie}$ is shifted down due to repulsion from
the bound state, whereas the influence of the quantizing magnetic
field on the level $E_{it_{2}}$ is negligible. As a result we
expect an increase of $\Delta_{CF}$ due to the Landau
quantization. A similar effect of the spatial quantization on the
crystal field splitting of TM impurities in semiconductor quantum
wells was discussed in \cite{KM95}.

\begin{figure}[htb]
\epsfysize=35 \baselineskip
\centerline{\hbox{\epsffile{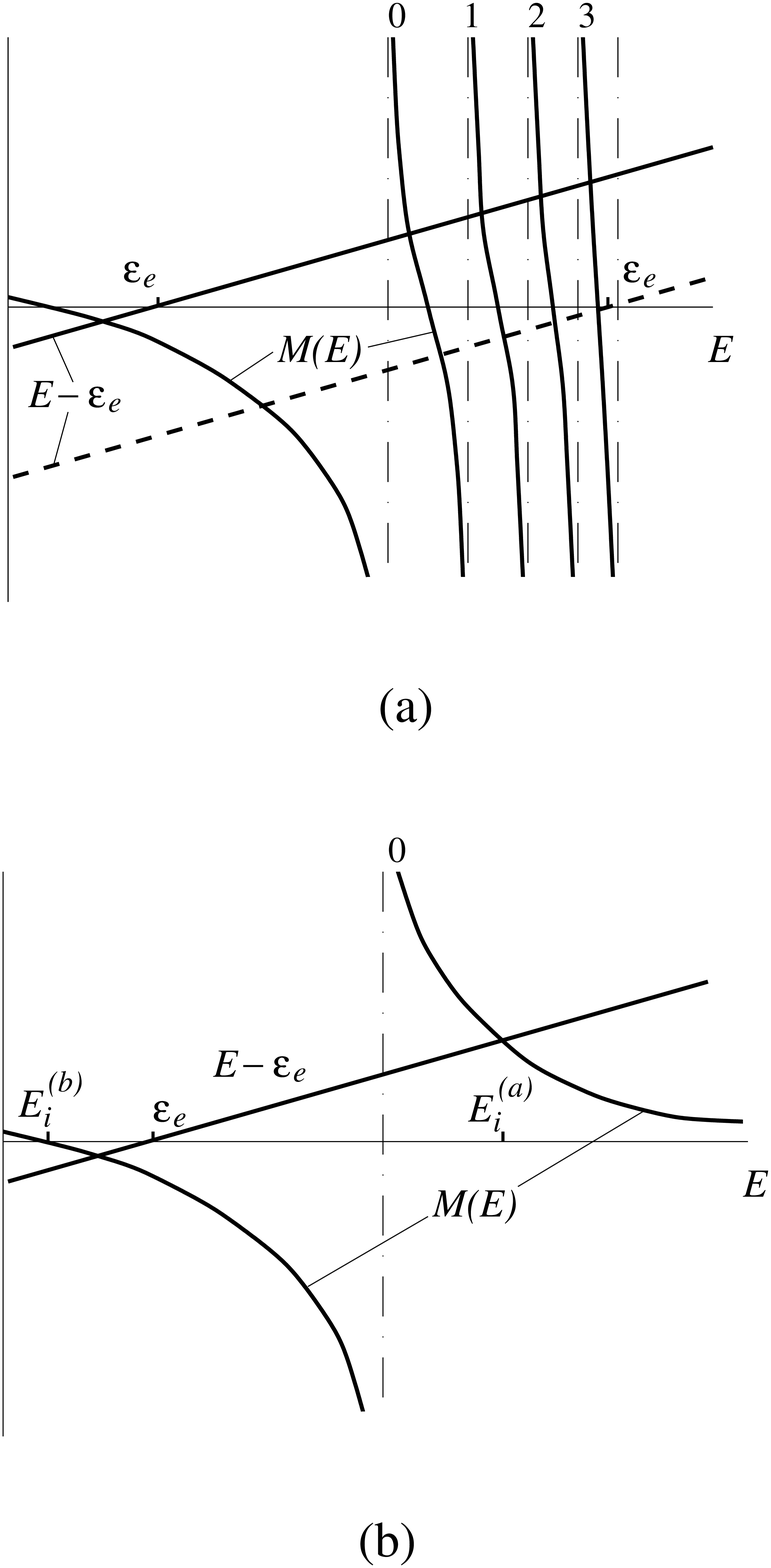}}}{\vspace{1cm}}
\caption{({\em a}) Graphical solution of Eq. (\ref{DM4}) for bound
impurity and Landau states in case of the $de$-level below the
quantized Landau grid (the solid $E - \varepsilon_e$ line) and in
case of the $de$-level within the Landau grid (the dashed $E -
\varepsilon_e$ line ). ({\em b}) Graphical solution of Eq.
(\ref{BA}) for the bound impurity and Landau states.}
\end{figure}

\bigskip

\begin{figure}[htb]
\epsfysize=19 \baselineskip
\centerline{\hbox{\epsffile{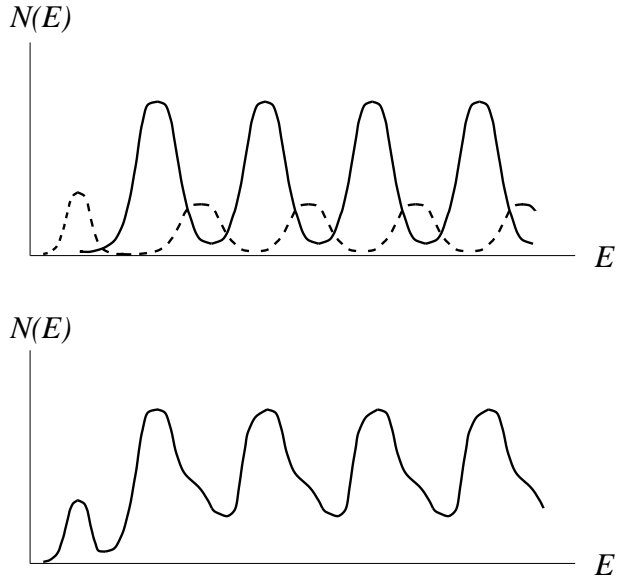}}{\vspace{1cm}}}
\caption{({\em a}) Density of free and bound Landau states
broadened by long range fluctuations of the local fields (solid
and dashed curves respectively. ({\em b}) Total density of states
of a doped 2DES.}
\end{figure}

\section{Spin structure of localized levels}

It is shown above that the resonance scattering does not change
radically the orbital part of the localized states formed by the
purely potential impurity scattering. However, taking into account
the spin structure of the impurity states results in more
significant distinctions between simple isoelectronic and TM
impurities in quantizing magnetic fields. In the latter case the
influence of TM impurity on Landau levels is {\em spin selective}.

To demonstrate this selectivity, one should take into account the
fact that the resonance states described by Eq. (\ref{IWF}) belong
to the impurity $3d$-shell with a definite configuration of
electron spins \cite{KF94}. Let us consider, for example, the
state of a TM impurity in a configuration d$^n$ where the last
($n$ - th) electron occupies the bonding level
$E_{ie\sigma}^{(b)}$ (\ref{BA}). Then the many-electron state of
the 3d shell may be represented as
\[
d^n = \left( ne_\uparrow^{r_1} e_\downarrow^{r_2}
t_{2\uparrow}^{r_3} t_{2\downarrow}^{r_4} \right)_{\sum_ir_i=n}.
\]
This notation means that the impurity has $n$ electrons in its
3$d$ shell, where $r_1$ and $r_2$ electrons with spin up and down,
respectively, are in the $e$-states, $r_3$ and $r_4$ electrons
with spin up and down, respectively, are in the $t_2$-states.

Normally, TM ions in a crystal field of III-V semiconductors exist
in the so called high spin state, which means that the
$t_{2\sigma}$ - and $e_{\sigma}$ - states are occupied in
accordance with the Hund rule. Therefore, the spins of the $e$
electrons in the $3d^n$ ions with $n \leq 5$ (from Ti to Mn) are
directed parallel to the external field ${\bf B}$. These electrons
form the deep energy levels $E_{ie\uparrow}$(d$^n$/d$^{n-1})$ well
below the bottom of the conduction band. The notation
(d$^n$/d$^{n-1})$, commonly accepted in the spectroscopy of deep
d-states in semiconductors \cite{Zung86a,KF94,Sokol87,LB}, means
that the occupation of the level $E_{ie\uparrow}$ corresponds to a
change of the atomic configuration from d$^{n-1}$ to d$^n$ due to
a transfer of a host spin-up electron to an e-state of the
impurity 3d shell. The levels $E_{ie\downarrow}$ are more shallow
than $E_{ie\uparrow}$ in accordance with the Hund rule, so that
$E_b - E_{ie\uparrow} \gg E_b - E_{ie\downarrow}$. Filling these
levels begins when the $3d$-shell is more than half-filled ($n>5
$, the elements from Fe to Ni). In this case both the
$E_{ie\uparrow}$ and $E_{ie\downarrow}$ levels are deep below the
bottom of the conduction band, and the resonance scattering is
weak in accordance with Eq. (\ref{BA}).

As a result, one can expect that the effect of the resonance
scattering will be strong for light elements (Ti, V, Cr, Mn), and
resonance interaction splits the $m=0$ states predominantly from
the down-spin Landau subband, whereas the potential scattering is
spin-independent. One can estimate the resulting spin splitting of
the lowest impurity Landau levels, $\Delta_{bs}$, from Eq.
(\ref{BA}):
\begin{equation}
\Delta_{bs} = \frac{|V_{eb}|^2} {\Delta_\uparrow \Delta_\downarrow}
\Delta_{es}
\label{SS}
\end{equation}
where $\Delta_{es} = \varepsilon_{e\downarrow} -
\varepsilon_{e\uparrow}$ is the exchange splitting of the impurity
d-levels, $\Delta_\sigma = E_b - \varepsilon_{e \sigma}$. The same
kind of spin splitting exists for the impurity levels belonging to
higher Landau bands. Fig. 3a illustrates the spin polarization of
Landau states in this case.

Hund rule is known to be violated for V impurity in some III-V
host crystals. Vanadium creates $e\sigma$-levels in the upper part
of the semiconductor energy gap for both spin projections. There
are numerous experimental \cite{Vasson93} and theoretical
\cite{Zung86} arguments in favor of the "anti-Hund" low-spin
states V$^0(e_{\uparrow} e_{\downarrow})$ and V$^-(e^2_{\uparrow}
e_{\downarrow})$ ions. Therefore, the resonance scattering is
strong for both Landau subbands and, moreover, the levels
$\varepsilon_{e\downarrow}$ and $E_b$ are nearly degenerate (Fig.
3b). In the case of a Cr ion the level $\varepsilon_{e\downarrow}$
is in resonance with the states above the bottom of the conduction
band in bulk GaAs, so the inequality $E_b <
\varepsilon_{e\downarrow}$ is valid (Fig. 3c). Then the spin
splittings of the Landau bound states and the impurity $de$ states
have the opposite signs (see next section). When calculating
$\Delta_{bs}$ in this case, one should use full Eq. (\ref{DM4})
instead of approximate Eq. (\ref{BA}).

Especially interesting is the case of Mn. Substitutional Mn ions
retain half filled d shell (Mn(d$^5$) with the maximal spin $S =
5/2$ according to the Hund rule) both in III-V and in II - VI
compounds. The corresponding levels $E_\gamma$(d$^5$/d$^4$) lie
deep in the valence band (see, e.g. \cite{Sz01}). This means that
the resonance scattering for spin up electrons is extremely weak.
The resonance level $E_{e\downarrow}$ corresponds to the empty
state Mn(d$^6$/d$^5$). This state was never observed directly, but
indirect data for some III - V \cite{Oka98} and II - VI compounds
\cite{Sokol87} indicate that such a level may exist within the
conduction band not far from its bottom. This means that Mn
impurity corresponds to the extreme limit of the case {\em (c)} in
Fig. 3. It should be mentioned that practically the same mechanism
of the magnetic coupling between TM ion and conduction electrons
via hybridization $V_{\gamma\mu,\lambda}$ (\ref{D1}) was
considered for the specific case of (Cd,Mn)Te/(Cd,Mn,Mg)Te
heterostructures in Ref. \onlinecite{Merk99}. Here the case of
zero magnetic field was considered, and $\lambda$ included
confined electrons near the edges of the valence and conduction
bands. It was shown in this paper that the "kinetic"
antiferromagnetic exchange $\sim |V_{\gamma\mu,\lambda}|^2/
(E_\lambda - E_{i\gamma})$ even in the case of a very deep
$d$-level is strong enough to compensate essential part of the
direct ferromagnetic exchange between the localized impurity spin
and the band carriers.

The spin-split Landau levels can be occupied by the electron-hole
pairs, and one can treat such pairs as the bound magnetic excitons
which appear on the background of conventional spin waves and
magnetoplasmons in 2DES. The theory of magnetic excitations bound
with 3$d$ impurities based on the methods of conventional theory
of magnetic excitations in 2DES \cite{BIE81,KH84,BMV96} will be
published elsewhere.

\begin{figure}[htb]
\epsfysize=16 \baselineskip
\centerline{\hbox{\epsffile{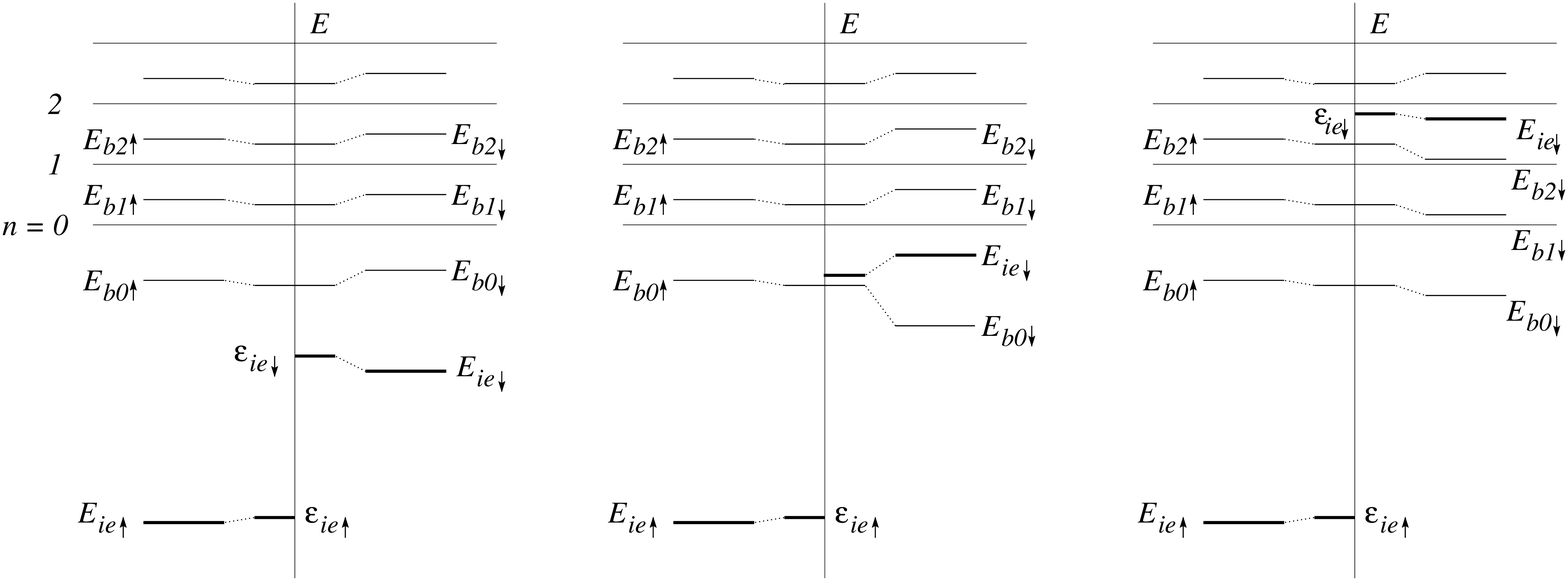}}} \caption{Energy levels
for the bound states : {(\em a)} weak scattering limit for both,
$\uparrow$ and $\downarrow$ states; {(\em b)} strong resonance
scattering for $\downarrow$ states; {(\em c)} the resonance level
for the $\downarrow$ states above the lowest Landau levels.}
\end{figure}

\section{Impurity $g$ factor}

During the last three decades the oscillatory enhancement of the
$g$ factor of delocalized electrons in 2DES was studied both
experimentally \cite {Fang68,Nicholas} and theoretically
\cite{Janak,Ando}. The oscillatory behavior of the $g$ factor in a
2DES is explained by a variation of the many-body Coulomb and
exchange renormalizations of the Zeeman splitting as a function of
the occupation of Landau subbands.

In this section we consider the behavior of the electronic $g$
factor due to the impurity scattering. As was shown in the
previous section, the spin splitting of local states with $m=0$ in
Landau subband may exists even in the absence of an external
magnetic field due to the spin selectivity of the resonance
impurity scattering. The mixed nature of this scattering potential
may cause a change of the $g$-factor provided the $g$-factors of
the impurity $d$-electrons and the quantized Landau electrons are
different. Since deviation of the $g$-factor from its bare
electron value $g_0 = 2.003$ may be noticeable enough both for
Landau electrons \cite{Nicholas} and for TM impurities \cite{LB}
in III-V semiconductors, the mutual influence of the two Zeeman
splittings may be also noticeable.

Three examples of spin splitting illustrated by Fig. 3 show the
variety of possible modifications of the $g$ factor as well. Let
us start with considering the Zeeman splitting of the electrons in
the lowest localized Landau state (weak scattering limit of Fig.
3a). In this case the deep d-levels $E_{i\uparrow,
\downarrow}^{(b)}$ with both spin projections are occupied and the
effective $g$ factor of the impurity is predetermined by the
Zeeman shift of the antibonding levels $E_{i\sigma}^{(b)}$. It may
be calculated by means of the simplified Eq. (\ref{BA}) or its
corollary (\ref{SS}). We define the effective $g$ factor in the
usual way,
\begin{equation}
\left\langle \hat{\psi}_{i}|H_{z}|\hat{\psi}_{i}\right\rangle = \pm
\frac{1}{2}\mu _{0}B_{\Vert }g_{eff},
\label{GFD}
\end{equation}
where $\hat{\psi}_i$ is the eigenvector with the components
$\psi_{i\sigma}^{(b,a)}$ which correspond to the solutions
(\ref{BA}) of the effective two-level problem.
\begin{eqnarray}
\psi _{i\sigma }^b & = &\cos \theta _\sigma \psi _{ie\sigma } + \sin
\theta _{\sigma }\psi _{b\sigma }  \nonumber \\
\psi _{i\sigma }^a & = &-\cos \theta _\sigma \psi _{b\sigma } + \sin
\theta _\sigma \psi _{ie\sigma }
\label{SR}
\end{eqnarray}
with the mixing coefficient given by
\[
\tan 2\theta _\sigma =\frac{2V_{eb}}{\Delta _\sigma }.
\]
$H_{z}$ is the Zeeman Hamiltonian
\begin{equation}
H_{z}=\mu _{0}\left( K{\bf L+}g_0{\bf S}\right) {\bf B},
\label{ZEE}
\end{equation}
{\bf S} and {\bf L} are the spin and orbital angular moments, $K$
is the covalency reduction factor. In our specific case of the
bound s-states the orbital contribution is absent. Extracting from
(\ref{GFD}) the antibonding component $\psi _{i\sigma}^a$, we
find, in the limit
\begin{equation}
\sin \theta_\sigma \approx V_{eb}/\Delta _{\sigma }\ll 1
\label{WEAK}
\end{equation}
and in the linear approximation in the magnetic field, that
\begin{equation}
\delta g =g_{eff} - g_b \approx - \frac{|V_{eb}|^2}{\Delta_{\uparrow}
\Delta _\downarrow} \left[\bar g\left(1 + \frac{\Delta _{es}^2}{2 \Delta
_{\uparrow}\Delta _\downarrow}\right) + \Delta _{es}L(|V_{eb}|^2)\right].
\label{ENHA}
\end{equation}
Here $\bar g = g_ b - g_ d $ and $g_{b,d}$ are the $g$ factors of
the Landau electron and the d$^n$ ion, respectively,
\[
g_d =g_s + g_L + \Delta g,
\]
$g_s$ and $g_L$ are the contributions to the $g$-factor due to the
projections of the spin and orbital moments on the total angular
moment. $\Delta g$ contains contributions due to the spin-orbit
interaction, electron-phonon interaction, etc. These corrections
are specific for a given ion in a given host semiconductor (see,
e.g., \cite{KF94}). Both the value and the sign of $\bar g$ may
vary in a wide range. $L(A) = A^{-1} \partial A/ \partial h$ is
the logarithmic derivative, $h=\mu _{0}B$.

It is seen from Eq. (\ref{ENHA}) that there are two contributions
into the enhancement of the $g$ factor. The first one is an
admixture of the polarization of the $d$ shell of the TM impurity
to the spin splitting of Landau states via the resonance
scattering channel. The second contribution to the enhancement
mechanism stems from the orbital effect: the hybridization matrix
element (\ref{int}) increases with the growing magnetic field
since the wavefunction $\psi_b(\xi ;E_{bn})$ becomes stronger
localized when the magnetic length decreases. Therefore, in the
weak scattering limit both contributions to the effective $g$
factor are positive provided $\bar g < 0$, and in this case we
deal with the enhancement (see below).

The renormalization of the $g$ factor of the bound Landau
electrons is more pronounced in the case when the Landau level
$E_{b0}$ and the impurity level $\varepsilon_{ie\downarrow}$ are
nearly degenerate (Fig. 3b). Then, confining ourselves with the
two-level approximation (\ref{BA}), we find that the inequality
$\sin\theta_\uparrow\ll 1$ is valid for the spin-up states,
whereas the mixing is strong for the spin-down states, so that
$\sin\theta_\downarrow \approx 1/\sqrt{2}$.  Neglecting the
orbital contribution we find that
\begin{equation}
\delta g_b =-\frac{\bar g}{4}\left( 1 - \frac{|\Delta _\downarrow
|} {2|V_{eb}|}+\frac{2|V_{eb}|^2}{\Delta _\uparrow^2}\right).
\label{ENHB}
\end{equation}
Here $-\bar g/4$ makes the principal contribution, and the two
other terms describe hybridization corrections for the spin-down
and -up states.

As expected, the sign and the magnitude of the $g$ factor
renormalization is predetermined by the difference $\bar g$. The
$g$ factor of the Landau electrons in the 2DES formed in GaAs is
small $g_b\ll g_0$ where $g_0=2.003$. On the other hand, the $g$
factor of 3d-ions varies from one element to another. It also
depends on the charge state of the given impurity and may be
influenced by local defects in the host crystal \cite{KF94,LB}.
However, nearly in all states its value is close to $g_0$. In
particular, $g_d=1.957$ for V$^{3+}$,  $g_d=1.974$ for Cr$^{3+}$
in bulk GaAs. The value of $g_d\approx 1.60$ was reported for
V$^{2+}$ \cite{Vasson84}, but this reduction is due to the
orthorhombic distortion of crystalline environment in this
specific sample. So we can assume in our estimates that $\bar g
\sim -1$ in the cases of our interest, and the $g$ factor
renormalization is, in fact, an {\em enhancement}. To estimate the
corrections to the maximum value of $\delta g_b =-\bar g/4$ given
by Eq. (\ref{ENHB}) one should notice that all three gaps
$\Delta_\uparrow,\ \Delta_\downarrow,\ \Delta_{es}$ are of the
same order in the range 0.5 to 1 eV in the weak scattering case
(\ref{WEAK}), and the hybridization parameter normally does not
exceed 0.1 to 0.2 eV.

The last term in the brackets of Eq. (\ref{ENHA}) can be evaluated
by means of Eq. (\ref{BWF2}) which is valid in the limit $\xi\gg
1$ (the magnetic length is essentially larger than the radius of
the d-wave function). A decrease of of the relative weight of the
Bloch tail of the s-type wavefunction $\psi_b$ implies a
corresponding growth of its "core" part, so that the resulting
increase of the logarithmic derivative $L(|V_{eb}|^2)$ can be
estimated as a small effect of the order of $(l_d/l_b)^2\ll 1$,
where $l_d\sim a_b$ is the radius of the $d$-part of the impurity
wave function (\ref{IWF}). Thus, the orbital contribution to the
Zeeman splitting is small in comparison with the paramagnetic one.
We conclude that even in this weak scattering limit the magnitude
of enhancement can be estimated as an effect $\sim (V_{eb}/\Delta
_{\sigma})^2$ which is up to 10\% above $g_0$.

In case of partially occupied Landau bands, an experimental
observation of a contribution of the b-levels with $m=0$, split
off the Landau bands $E_n\neq 0$, on the background of other
states filling the gaps between Landau levels (Fig. 2) may be not
an easy task. At small TM impurity concentrations the strongest
observable effect involving Zeeman splitting is a variation of the
impurity d-shell occupation in a quantizing magnetic field. The
hybrid structure of the impurity states can be detected when
studying the d-electron $g$ factor, e.g., by methods of ESR
spectroscopy, provided the impurity configuration is
e$^1$b$^{2n}$, i.e., the level  $E_{i\downarrow}$ is empty in the
zero magnetic field. Noticeable effects are expected when
$E_{i\downarrow}$ is slightly above the Fermi level
$\varepsilon_F$ of the conduction electrons in the zero magnetic
field. (Fig. 4).

\begin{figure}[htb]
\epsfysize=20 \baselineskip
\centerline{\hbox{\epsffile{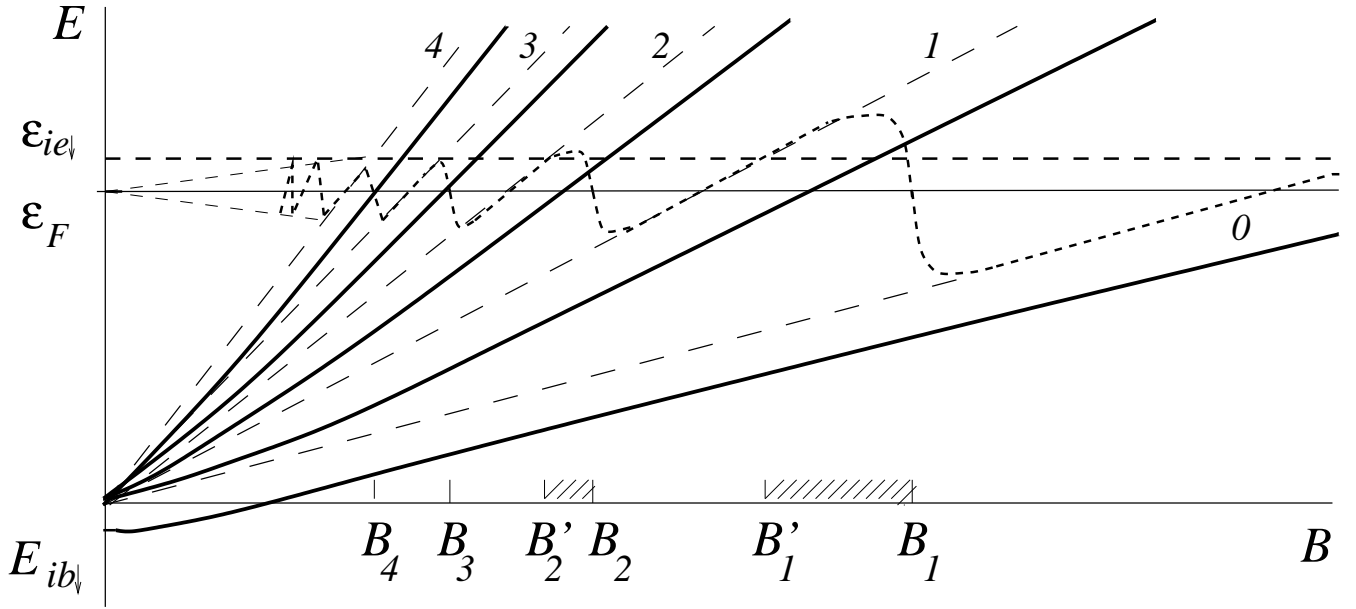}}}\vspace{.5cm}
\caption{Relative positions of bound Landau states
$E_{bn\downarrow}$ (thick solid lines), free Landau levels $E_n$
(thin dashed lines), resonant level $\varepsilon_{ie\downarrow}$
(thick dashed line) and the Fermi level $\varepsilon_F$ in the
zero magnetic field (thin line) and in a quantizing magnetic field
(thick dotted line). See text for further explanations.}
\end{figure}

The dotted saw-tooth-like curve in this figure shows the
oscillations of the Fermi level $\varepsilon_F(B)$ that reflect
the jumps $\delta_n$ from the filled Landau level $E_n$ to the
next Landau level $E_{n+1}$ at $B=B_n$. These jumps $\delta_n =
\hbar^2/ml_{B_n}^2$ are smoothed, since the inter-level windows
are filled with impurity states of various origin (see Fig. 2).
Since the diamagnetic shift of the Landau levels with an
increasing magnetic field becomes essentially bigger than the
Zeeman splitting $(\hbar^2/2m^*l_B^2\gg g\mu_B B)$, the latter
effect can be neglected in this evolution of the spectrum. Then
with increasing field $B$, the impurity e-level will cross all
Landau levels, and eventually, at a high enough magnetic field, it
will be squeezed out into the energy gap below the first Landau
level. If the difference $E_{i\downarrow} - \varepsilon_F <
\delta_n$, then the occupation of the d-shell changes from
e$^1$b$^{2n}$ to e$^2$b$^{2n}$ within some intervals $B_1^\prime
\div B_1$ and $B_2^\prime\div B_2$ of the magnetic field (hatched
domains in Fig.4), and the ESR signal should disappear in these
regions. Small dips in $g_d$ due to an admixture of the b-states
may be also observed near the points $B_3$, $B_4$, etc.

Recent achievements in fabricating heterostructures with large
manganese contents, e.g., (In, Mn)As/(Al, Ga)Sb, (Cd, Mn)Te/(Cd,
Zn, Mg)Te [see, e.g., \cite{Dietl}] open a new possibility of
tuning the electron $g$-factor. Certainly, one should be cautious
when applying our theory, formulated for isolated TM impurities,
to systems with high impurity concentrations. In particular, an
interaction between impurities may take place leading to a
concentration dependence of the $g$ factor \cite{Potem00}. Anyhow,
the number of electron in bound Landau states is noticeable in
this case. One can expect that the impurity $g$ factor will change
in accordance with equations (\ref{ENHA}) and (\ref{ENHB}) each
time when the level $\varepsilon_{ie\downarrow}$ crosses a bound
Landau level $\varepsilon_{bn\downarrow}$. Probably, the minor
oscillations of the $g$-factor, observed on the background of the
strong $g$-factor change due to the above concentration effects in
magneto-optical spectra of (Cd, Mn)Te/(Cd, Mg)Te \cite{Potem99}
heterostructures, can be explained by the mechanism proposed in
this section. These oscillations will be a subject of our future
investigation.

\section{Conclusions}

This paper demonstrates that the Landau quantization makes
dramatic changes in the structure of localized states in
semiconductors doped by magnetic impurities. We concentrate here
on the appearance of bound Landau states near the bottom of the
conduction band. It is found that both orbital and spin parts of
the impurity scattering potential are significant. We consider TM
ions substituting cations in III-V semiconductors and have found
that both short-range potential and resonance components of
impurity scattering influence mainly the states with the zero
orbital momentum $m=0$.

New features introduced by the resonance component of the
scattering potential, as compared to the short range potential
considered in \cite{Avishai93,Avishai97}, can be summarized as
follows: ({\em i}) the sign of the scattering potential depends on
the position of the resonance level $E_{ie}$: the resonance
scattering partially cancels the short-range potential scattering,
when $E_{ie}-E_b <0$, and enhances it, when $E_{ie}-E_b >0$; ({\em
ii}) the resonance scattering plays a decisive part in appearance
of the charged bound states.

The role of spin effects is more profound: the fact that the spin
state of a TM ion is determined by the interatomic Coulomb and
exchange interaction, makes the resonance scattering spin
selective. Its influence on the Landau states also depends on the
spin state of the Landau orbitals. The most striking manifestation
of this spin selectivity is a possibility of a significant
enhancement of the $g$ factor of the localized Landau states. This
enhancement may be noticed in magneto-optical measurements.

The same resonance scattering results also in noticeable changes
of the properties of the electronic states of TM impurities: ({\em
i}) removal of the symmetry ban for hybridization between the
de-states and s-states for the conduction band may result in a
considerable increase of the crystal field splitting
$\Delta_{CF}$; ({\em ii}) reduction of the $g$ factor of the
d-electrons is dual to that of the corresponding Landau states, so
that an experimental possibility arises to study properties of
bound states in Landau bands by measuring the $g$ factor of the
impurity $d$-shell. It is worth noting that the reduction of the
$g$ factor may be responsible for an enhancement of the nuclear
spin relaxation, because it partially compensates the difference
in the energy scales of nuclear and electronic subsystems.

The modification of the one-electron spectrum of a 2DES by the
impurity scattering should also be accompanied by changes in the
spectrum of spin excitons, magnetoplasmons and other collective
excitations in strongly quantized electron gas. These effects will
be described in forthcoming publications.

{\bf Acknowledgement} This work was supported by the German -
Israeli Foundation for Research and Development, Grant
No.0456-220.07195. K.K. thanks Israeli Science Foundation for
support (grant "Nonlinear Current Response of Multilevel Quantum
Systems"). The authors are indebted to S. Dickmann, S. Gredeskul,
and M. Potemski for valuable discussions. VF and KK are grateful
to Max Planck Institute for Complex Systems, Dresden, for
hospitality.

\end{document}